\documentclass[aps,prb,preprint,amssymb,superscriptaddress,showpacs]{revtex4-1}

\usepackage{verbatim}
\usepackage[english]{babel}
\usepackage{amsmath,amsthm}
\usepackage{graphicx}
\usepackage{subfigure}
\usepackage{epsfig}
\usepackage{threeparttablex}
\usepackage{natbib}
\renewcommand{\[}{\left[}
\renewcommand{\]}{\right]}
\renewcommand{\(}{\left(}
\renewcommand{\)}{\right)}

\newcommand{\vect}[1]{\mathbf{#1}}
\newcommand{\tens}[1]{\underline{\underline{#1}}}

\begin{document}
\title{Pseudo--potentials based first--principles approach to the magneto--optical
       Kerr effect: from metals to the inclusion of local fields and excitonic effects}

\author{Davide Sangalli}
\affiliation{MDM Lab - IMM - CNR via C. Olivetti, 2 I-20864 Agrate Brianza (MB) Italy, European Union}
\affiliation{European Theoretical Spectroscopy Facilities (ETSF)}

\author{Andrea Marini}
\affiliation{Istituto di Struttura della Materia of the National Research Council, Via Salaria Km 29.3,
I-00016 Monterotondo Stazione, Italy}

\author{Alberto Debernardi}
\affiliation{MDM Lab - IMM - CNR via C. Olivetti, 2 I-20864 Agrate Brianza (MB) Italy, European Union}

\date{\today}

\begin{abstract}
We propose a first--principles scheme for the description of the magneto-optical kerr effect
within density functional theory (DFT). 
Though the computation of Kerr parameters is often done within DFT, starting from the
conductivity or the dielectric tensor, there is no formal justification to this choice.

As a first steps, using as reference materials iron, cobalt and nickel we show
that pseudo--potential based calculations give accurate predictions.
Then we derive a formal expression for the full dielectric 
tensor in terms of the density--density correlation function.
The derived equation is exact in systems with an electronic gap, with
the possible exception of Chern insulators, and whenever
the time reversal symmetry holds and can be used as a starting point for the
inclusion of local fields and excitonic effects within time--dependent
DFT for such systems.

In case of metals instead we show that, starting from the density--density
correlation function, the term which describes the anomalous Hall effect is
neglected giving a wrong conductivity.
\end{abstract}

\maketitle

\section{Introduction}
The Magneto--Optical Kerr Effect\,(MOKE) consists in the rotation of the polarization plane of light
reflected from the surface of a magnetic material. It was discovered in 1877 by John Kerr~\cite{Kerr1877,Kerr1878}
while he was examining the light reflected from a polished electromagnet pole.
Very recently it became the object  of an intense experimental investigations, mainly for two
reasons. First one can exploit this effect to read suitably magnetically stored information using
optical means in modern high density data storage technology~\cite{Bertero1994,Hatwar1997}.
Second, the MOKE can be used as a powerful probe in many fields of
research such as microscopy for domain observation, surface magnetism, magnetic interlayer coupling
in multilayers~\cite{Hatwar1997,Bennett1990,Suzuki1998,Zinoni2011,Balk2011}.
It can also be used to observe plasma resonance effects in thin layers,
and structural and magnetic anisotropies~\cite{Wu1999,Zeper1989,Weller1992}.

The microscopic origin of the Kerr effect is a combined action of the spin-orbit coupling (SOC)
and the net spin--polarization of the material~\cite{Guo1995}.
Indeed the existence of a non zero magnetization in the ground state is due to 
a spontaneous symmetry breaking of the system.
This symmetry breaking is transferred, through the SOC, to the 
spatial part of the wave--functions so that $\psi_{+L_z}(\mathbf{x})$ is different from $\psi_{-L_z}(\mathbf{x})$.
Accordingly the absorption is different for light with right and left circular polarization.

The problem has been addressed in the literature and {\it ab--initio} calculations, based on
density functional theory (DFT),
are available for transition metals like $Fe$, $Co$ and
$Ni$~\cite{Delin1999,Oppeneer1992,Oppeneer1995,Guo1995,Kim1999,Luttinger1967,Vernes2002,Vernes2004A,Vernes2004B}
and, recently, for other materials such as
full-Heusler films and $Mn$--doped $GaAs$~\cite{Ricci2007,Stroppa2008,Uba2012,Ricci2011, Haidu2011}.

The Kerr parameters are commonly obtained from the Kubo formula~\cite{Kubo1957}
for the optical conductivity tensor using the single particle Kohn--Sham (KS) wavefunctions.
This is equivalent to the computation of the dielectric constant at
the random phase approximation (RPA) without the 
inclusion of local fields (LF) and exchange--correlation ($xc$) corrections.
We will refer to this approach as independent particles RPA (IP--RPA) scheme.

Also an alternative approach, based on the Luttinger's Formula~\cite{Luttinger1967}, has been 
proposed in the literature~\cite{Vernes2002,Vernes2004A,Vernes2004B},
starting from the current--current correlation function, $\tens{\chi}_{\mathbf{jj}}$.
Also in this case the KS wave--functions are used and LF and xc--effects
are not considered.

In all these works the Kerr paramenters are computed within the DFT framework strarting
from the dielectric tensor, $\tens{\varepsilon}(\omega)$, or, which is the same, from
the conductivity, $\tens{\sigma}(\omega)$ .
However while the diagonal terms of the dielectric tensor,
i.e. $\varepsilon_{ii}(\omega)$, can be expressed in terms of the density--density correlation
function, $\chi^0_{\rho\rho}$ (or $\chi_{\rho\rho}$ if LF and xc effects are included),
at the best of our knowledge, no such expression has
been derived for the off--diagonal terms, i.e. $\varepsilon_{ij}(\omega)$ with $i\neq j$.
The latter can be obtained only starting from $\tens{\chi}^0_{\mathbf{jj}}$
which however is not expressed as a functional of the density.
In this case the sole formal justification to the use of KS quantities to construct
$\tens{\chi}^0_{\mathbf{jj}}$ is that ``a posteriori'' the approach give good
results and that KS wave--functions can be regarded as a good approximation
to quasi--particle wave--functions. 

In the present work instead we derive an expression for the full dielectric
tensor in terms of $\chi^0_{\rho\rho}$ and so for the construction of the
Kerr parameters within a density based approach. Besides a formal justification
to the use of the DFT approach, the result we propose can be regarded as a starting point
to go beyond the RPA--IP scheme to include LF and xc--effects.
Indeed while the description of transition metals within the RPA--IP approach is reasonable,
for semiconductors important deviations are expected, in particular when excitons
(magnetic excitons) exist. Magnetic semiconductors are in fact materials of great interest,
especially in view of spin electronics (spintronics) applications and
the MOKE can be a valuable tool for the investigation
of their properties~\cite{Acbas2009,Sun2011}.

In the present work we limit our discussion to the polar geometry, that is when the propagation direction
of the photon (the $z$ axis) and the magnetization of the system are both perpendicular
to the surface of the sample ($xy$). Experimentally this is the most studied geometry,
and it is also the one which in general gives the largest MOKE signal.
To support our theoretical derivation with numerical results we
have implemented in the plane--waves and pseudo--potentials based
code $Yambo$~\cite{Yambo} the computation of the Kerr parameters.
We have tested our implementation in transition metals for which well
assessed all electrons calculations and experimental results are available.
For these materials the IP--RPA approximation is sufficient.

Thus in Sec.~\ref{sec:MOKE_IP_RPA} we show results on bulk iron, cobalt and nickel in order
to validate our pseudo--potential based approach, at the IP--RPA level.

Then we discuss how to construct a formal equation for the dielectric tensor 
starting from $\chi^0_{\rho\rho}$ in Sec.~\ref{sec:MOKE_FULL}.
This approach is compared with the one based on $\tens{\chi}^0_{\mathbf{jj}}$.
The two differ by a term which, in general,  is zero in systems with an electronic
gap or whenever the pure time reversal symmetry holds.
This term describes the anomalous Hall conductivity.
Taking iron as reference system we show that, neglecting this term,
a large error is induced in the computation of the off diagonal conductivity in metals.
However the anomalous Hall conductivity is zero in systems with an electronic gap~\cite{footnote_chern},
thus our approach is exact for dielectrics. We then show how LF and xc--effects can be included
replacing $\chi^0_{\rho\rho}$ with $\overline{\chi}_{\rho\rho}$.
The results is a scheme, in principle exact, to compute the Kerr parameters within
time--dependent DFT (TDDFT).

\section{Moke parameters with the IP--RPA approximation}\label{sec:MOKE_IP_RPA}
\subsection{Theoretical background}

The description of the MOKE can be obtained in terms of the dielectric function $\varepsilon(\omega)$
or equivalently of the optical conductivity $\sigma(\omega)$. The two are related by the
equation
\begin{equation}\label{eq:sigma_eps}
\tens{\sigma}(\omega)=\frac{\omega}{4 \pi i}(\tens{\varepsilon}(\omega)-\tens{1})
\text{,}
\end{equation}
where the dielectric tensor at the IP--RPA level can be constructed from the
(paramagnetic) $\chi^0_{\textbf{jj}}$, according to the equation~\cite{Strinati1988}:
\begin{equation}    \label{eq:eps_jj} 
\varepsilon_{\alpha,\beta}(\omega)=\(1-\frac{4\pi e^2n}{m\omega^2}\)\delta_{\alpha,\beta}-
                             \frac{4\pi e^2}{\omega^2}\ \chi^0_{j_\alpha j_\beta}(\mathbf{0},\omega)
\text{.}
\end{equation}
Here $e$ is the electron charge and $m$ the electron mass, $\Omega$ the unit cell volume,
$\alpha$ labels the Cartesian axis, 
and $n=N_{el}/\Omega$ the number of electrons per unit volume.
The IP response functions is:
\begin{eqnarray}\label{eq:dipole_psp}
\chi^0_{j_\alpha j_\beta}(\mathbf{q},\omega)&=&\sum_{cv} \int \frac{d^3\mathbf{k}}{(2\pi)^3}
      \[\chi^0_{j_\alpha j_\beta}(\mathbf{k},\omega)\]_{cv\mathbf{kq}}
       \nonumber \\
       &=&\frac{1}{m^2} \sum_{cv} \int \frac{d^3\mathbf{k}}{(2\pi)^3}  \left [ 
        \frac{(p_{cv\mathbf{k}}^\alpha(\mathbf{q}))^*  p_{cv\mathbf{k}}^\beta(\mathbf{q})\ f_{v\mathbf{k}}(1-f_{c\mathbf{k-q}})}
             {\hbar\omega-(\epsilon_{c\mathbf{k-q}}-\epsilon_{v\mathbf{k}})+i\eta} - \right .
       \nonumber \\
&&\phantom{ \sum_{cv} \int \frac{d^3\mathbf{k}}{(2\pi)^3} \[\chi^0_{j_\alpha j_\beta}(\mathbf{q},\omega)\]_{cv\mathbf{kq}}} \left .
        \frac{(p_{vc\mathbf{k}}^\alpha(\mathbf{q}))^*  p_{vc\mathbf{k}}^\beta(\mathbf{q})\ f_{v\mathbf{k-q}}(1-f_{c\mathbf{k}})}
{\hbar \omega+(\epsilon_{c\mathbf{k}}-\epsilon_{v\mathbf{k-q}})+i\eta} \right ],
\end{eqnarray}
where the first term on the r.h.s. is the resonant part, while the second in the anti-resonant one.
$c$ ($v$) are conduction (valence) band indexes, $f_{i\mathbf{k}}$ are the occupation factors,  $\epsilon_i({\bf k})$
the electronic energy at ${\bf k}$ and
$p_{cv\mathbf{k}}^\alpha(\mathbf{q})$ the expectation value of the momentum operator
${\langle c\mathbf{k-q} |\hat{p}^\alpha| v\mathbf{k} \rangle}$ which in our pseudo--potential
based scheme must be computed as~\cite{Read1991}
\begin{equation}
\langle c\mathbf{k-q} |\left(\hat{p}^\alpha
-\frac{im}{\hbar}[\hat{x}^\alpha,\hat{V}_{NL}]\right)| v\mathbf{k} \rangle
\text{,}
\end{equation}
where $\hat{x}^\alpha$ is the $\alpha$ component of the position operator and
$V_{NL}$ the non--local part of the pseudo--potential. The infinitesimal $\eta$ factor
implies that the electromagnetic field is adiabatically turned on at $t=-\infty$ but can also be viewed as 
a finite lifetime broadening which accounts for scattering process and finite experimental resolution.
In the present work we use $\eta(\omega)=0.3\ eV+0.03\hbar\omega $ to mimic an experimental
resolution which decrease linearly with energy as in Ref.~\onlinecite{Delin1999}.

\subsection{MOKE spectra for transition metals}\label{sec:results_IP}

As a first step we check how a pseudo--potentials based approach preforms
in the description of the Kerr parameters, as it is common wisdom~\cite{Delin1999}
that all electrons calculation are needed to describe the wave--function
in the core region, where the SOC is mostly effective, and thus to evaluate the 
MOKE.

We start from a ground state DFT calculation for bulk iron, cobalt and nickel
with the $Abinit$ code~\cite{Abinit},
using norm--conserving HGH~\cite{HGH} pseudo--potentials and including the SOC;
we found out that it is crucial, for a correct description of the MOKE,
to have the SOC effect included both in the pseudo--Hamiltonian and in the
construction of the pseudo--potential.

\begin{figure}[!ht]
\includegraphics[width=0.75\textwidth]{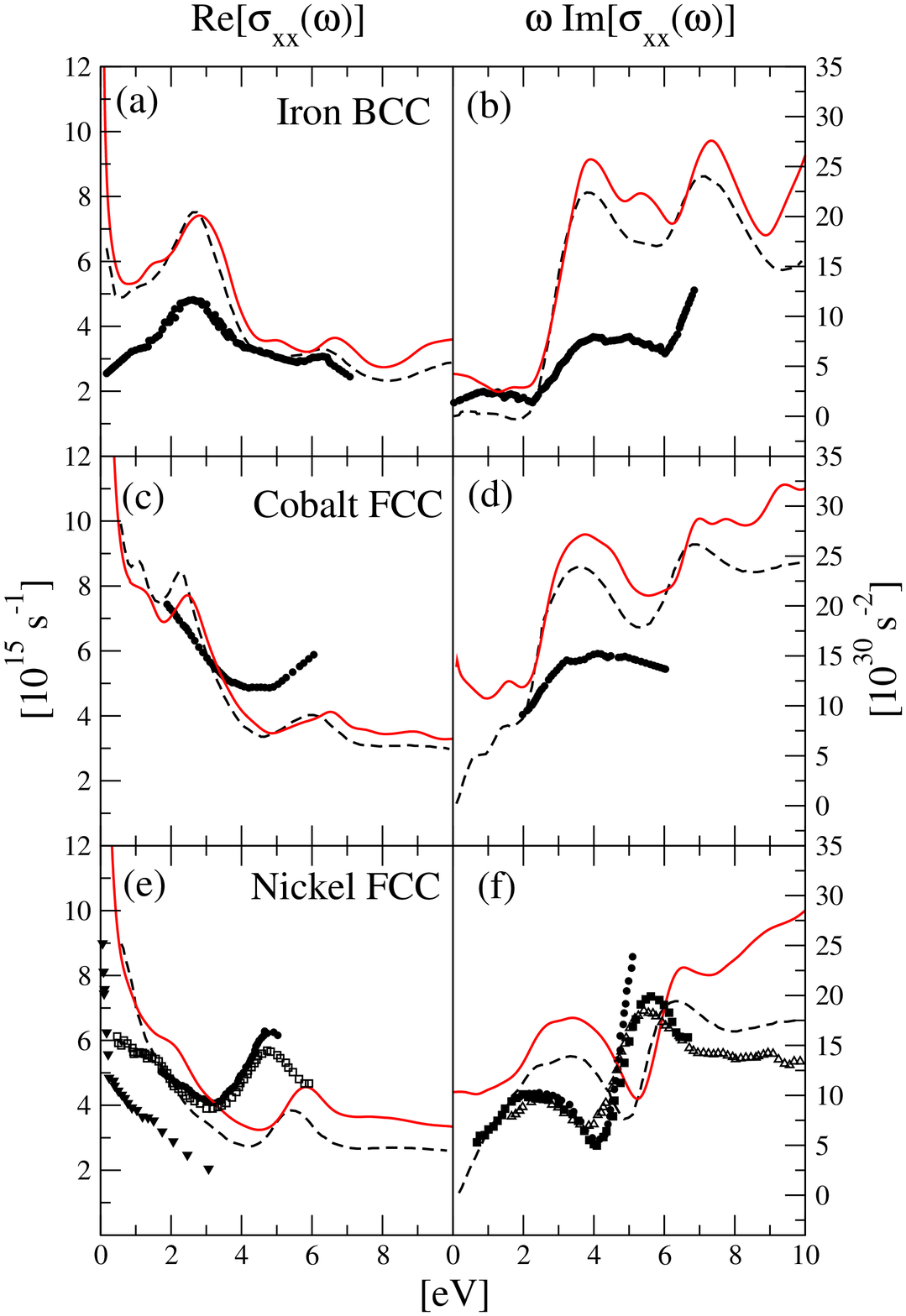}
\caption{(color online) Plot of $\sigma_{xx}(\omega)$ for bulk BCC iron (Panels $a$-$b$),
bulk fcc cobalt (Panels $c$-$d$) and bulk fcc nickel (Panels $a$-$b$). 
The continuous (red) line are the results from the preset work. The dashed line
are all electrons results from Ref.~\onlinecite{Delin1999}.
The symbols are experimental measurements.
Panels $a$-$b$: Ref.~\onlinecite{Johnson1974}.
Panels $c$-$d$: Ref.~\onlinecite{Nakajima1996}.
Panels $e$-$f$: filled circels, Ref.~\onlinecite{Nakajima1996}; empty squares, Ref.~\onlinecite{Shiga1969};
                filled triangles, Ref.~\onlinecite{Stoll1971}; filled squares, Ref.~\onlinecite{Johnson1974};
                empty triangles, Ref.~\onlinecite{Ehereinreich1963}.}
\label{fig:Sigma_xx}
\end{figure}

\begin{figure}[!ht]
\includegraphics[width=0.75\textwidth]{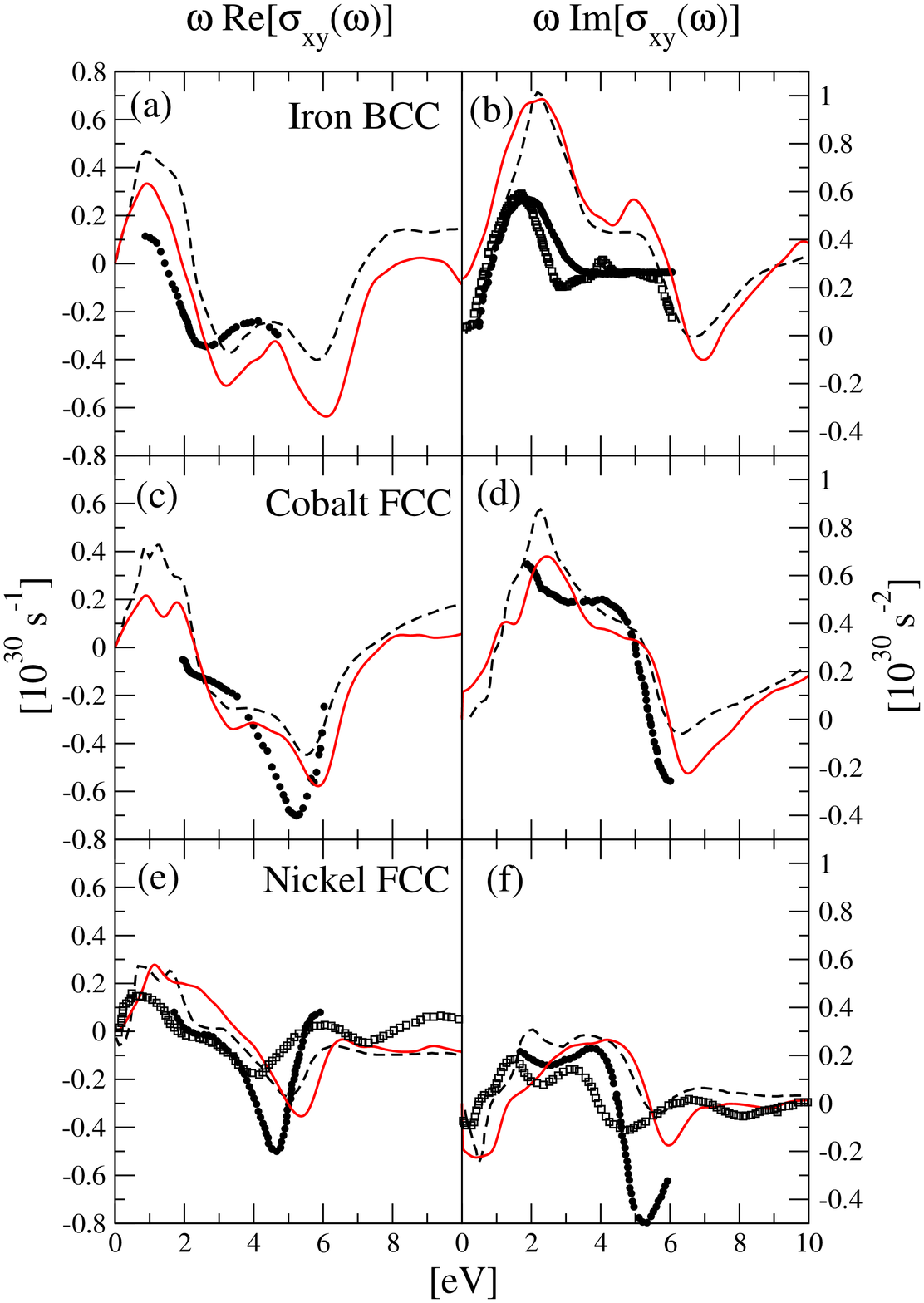}
\caption{(color online) Plot of $\sigma_{xy}(\omega)$ for bulk BCC iron (Panels $a$-$b$),
bulk fcc cobalt (Panels $c$-$d$) and bulk fcc nickel (Panels $a$-$b$) 
The continuous (red) line are the results from the preset work. The dashed line
are all electrons results from Ref.~\onlinecite{Delin1999}.
The symbols are experimental measurements.
Panels $a$-$b$: filled circles, Ref.~\onlinecite{Engen1983};
                empty squares,  Ref.~\onlinecite{Krinchik1968}.
Panels $c$-$d$: Ref.~\onlinecite{Nakajima1996}.
Panels $e$-$f$: filled circels, Ref.~\onlinecite{Nakajima1996}; empty squares, Ref.~\onlinecite{Erskine1977}.}
\label{fig:Sigma_xy}
\end{figure}

Bulk iron is studied in its bcc phase with the experimental cell parameter
$a=2.87$ \AA; an energy cutoff of 65 $Ha$ and a k--points sampling of the Brillouin zone (BZ) 14x14x14.
Bulk cobalt is studied in its fcc phase with the experimental cell parameter
$a=3.55$ \AA; an energy cutoff of 55 $Ha$ and a k--points sampling of the BZ 8x8x8.
Finally bulk nickel is studied in its fcc phase with the experimental cell parameter
$a=3.52$ \AA; an energy cutoff of 65 $Ha$ and a k--points sampling of the BZ zone 14x14x14.
Semi--core electrons are also included in the pseudo--potentials for all systems,
as we found the density of states to be
poorly described using the HGH pseudo--potentials with only valence electrons, constructed from
the parameters of Ref.~\onlinecite{HGH}.

Then we compute the dielectric function with the $Yambo$ code~\cite{Yambo} according to a modified
version of Eq.~\ref{eq:eps_jj}:
\begin{equation}
\varepsilon_{\alpha,\beta}(\omega)=\(1+\frac{4\pi e^2}{\omega^2}\chi^0_{j_\alpha j_\beta}(\mathbf{0},0)\)
                             \delta_{\alpha,\beta}-
                             \frac{4\pi e^2}{\omega^2}\ \chi^0_{j_\alpha j_\beta}(\mathbf{0},\omega)
\text{.}
\end{equation}
where the diamagnetic term has been replaced by the zero frequency value of $\chi_{jj}$.
Indeed in cold semi--conductors the diamagnetic term must be exactly balanced by
$\chi^0_{j_\alpha j_\beta}(\mathbf{0},0)$,
due to the effective--mass sum rule~\cite{Sipe1993,Virk2007,Giuliani2005},
but this balance is slowly converging with the number of $cv$ states included
and the use of $\chi^0_{j_\alpha j_\beta}(\mathbf{0},0)$ speeds up the convergence~\cite{footnote_Drude}.
In metals instead the difference
between the diamagnetic term and the zero frequency value of $\chi_{jj}$ gives the Drude term,
which, with this choice, is set to zero. However it has been shown~\cite{Marini2001} that
in practice a ultra--fine sampling of the BZ would be needed to compute this difference.
Hence it is preferable to include it with a semi--classical model
as described in Ref.~\onlinecite{Marini2001}.
The Drude term is included only in the computation of the diagonal part of
$\tens{\varepsilon}(\omega)$.
The optical conductivity is finally constructed from Eq.~\ref{eq:sigma_eps}.
Here for the Drude term we used the same parameters of the reference
all--electrons calculation, i.e. ${\omega_P=(4.9+i1.8\pi)\ eV}$ for iron,
$\omega_P=(8.3+i2\pi)\ eV$ for cobalt and
$\omega_P=(7.5+i2.24\pi)\ eV$ for nickel.

The results for the optical conductivity are plotted in Figs.~\ref{fig:Sigma_xx}-\ref{fig:Sigma_xy}.
For all systems there is a systematic blue shift of the theoretical peaks against the experimental data.
This is a known problem of the LDA, due to the self--interaction error, which
tends to delocalize the $d$--orbitals and accordingly gives wrong eigenvalues.
The same consideration also explains the overestimation of the intensity, as delocalization 
increases the orbitals overlap and thus the intensity of the dipoles computed to construct the
dielectric function.
However a good agreement is found with the reference all--electrons calculations.
The diagonal component, $\sigma_{xx}(\omega)$, is commonly computed from pseudo--potential based
calculations, provided that the dipoles are constructed as in Eq.~\ref{eq:dipole_psp}.
For the off--diagonal component, $\sigma_{xy}(\omega)$, instead, it has been reported that it
must be computed using all--electrons wavefunctions~\cite{Delin1999}, because it
depends crucially on the correction to the wave--function due to the SOC term in the Hamiltonian.
However, in our results, the differences in the diagonal and the off--diagonal part of the optical
conductivity compared to the reference all--electrons calculations are of the same order.
Hence we can conclude that pseudo--potentials based calculation can be used to compute the Kerr
parameters with the same level of confidence of absorption spectra, which depend only on the diagonal
part of the optical conductivity.
We mention that, in the literature, pseudo--wavefunctions has been used to construct
the off--diagonal conductivity at $\omega=0$ in the computation of the anomalous Hall
conductivity~\cite{Yafet2007,Wang2006,Wang2007}.
Also in this case a good agreement with the all--electrons calculations was found.

\begin{figure}[!ht]
\includegraphics[width=0.75\textwidth]{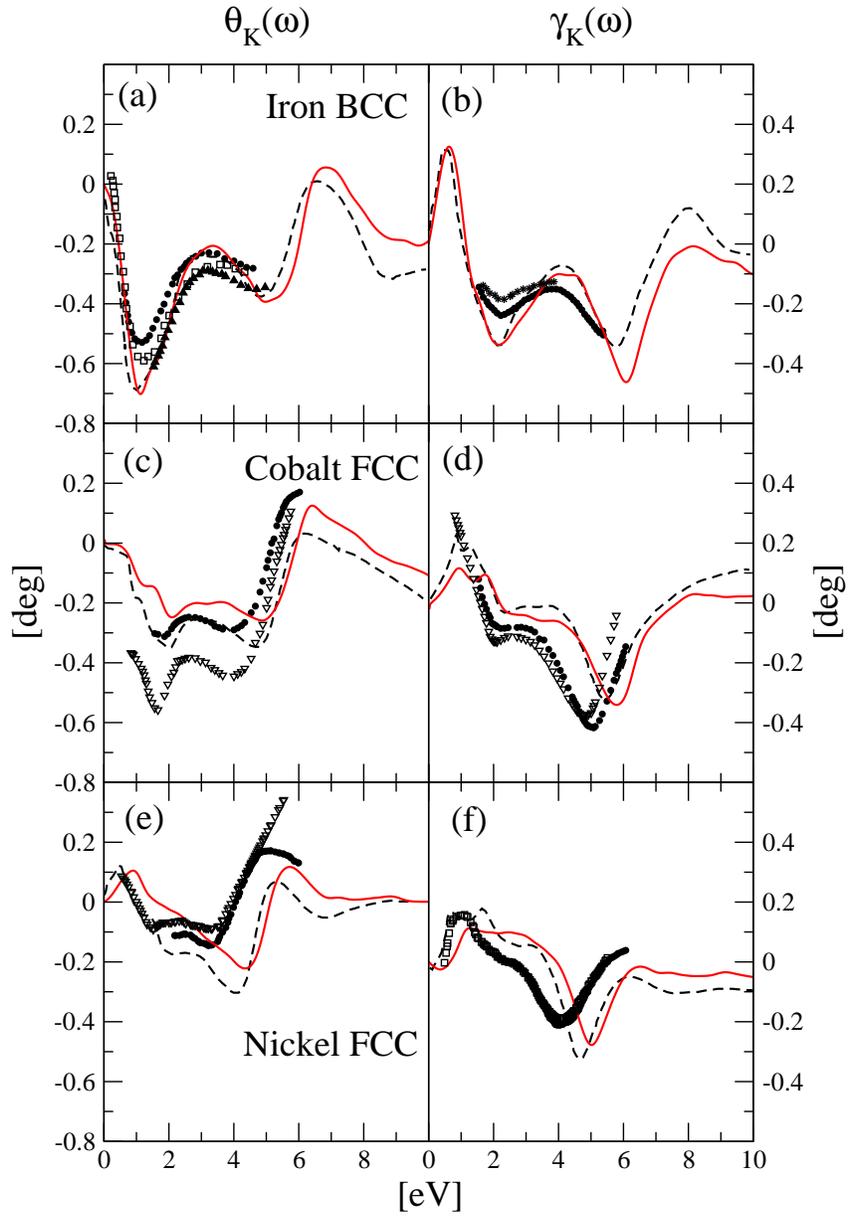}
\caption{(color online) Plot of the Kerr paramenters $\Psi_K(\omega)=\theta_K+i\gamma_K$
         for bulk BCC iron (Panels $a$-$b$), bulk fcc cobalt (Panels $c$-$d$) and bulk fcc nickel (Panels $a$-$b$).
The continuous (red) line are the results from the preset work. The dashed line are all electrons results from
Ref.~\onlinecite{Delin1999}. 
The symbols are experimental measurements.
Panels $a$-$b$: filled circles, Ref.~\onlinecite{Engen1983};
                empty diamonds, Ref.~\onlinecite{Krinchik1968}; empty triangles, Ref.~\onlinecite{Katayama1986};
                stars, Ref.~\onlinecite{Visnovsky1996}; filled squares, Ref.~\onlinecite{Kawagoe1992}.
Panels $c$-$d$: filled circels, Ref.~\onlinecite{Nakajima1996}; empty triangles, Ref.~\onlinecite{Weller1994}.
Panels $e$-$f$: filled circels, Ref.~\onlinecite{Nakajima1996}; empty triangles, Ref.~\onlinecite{Krinchik1968};
                Empty squares, Ref.~\onlinecite{Engen1983}.}
\label{fig:Kerr}
\end{figure}

Thus we finally compute the complex Kerr parameters according to the equation
\begin{equation}
\Psi_K(\omega)=\theta_K(\omega)+i\gamma_K(\omega)=\frac{-\varepsilon_{xy}}{(\varepsilon_{xx}-1)\sqrt{\varepsilon_{xx}}}
\text{,}
\end{equation}
which is the standard expression for the polar geometry in the small angles limit. Here
the photon propagates along the $z$ direction and describes a linearly polarized wave with the
electric field along the $x$ direction. Results are reported in Fig.~\ref{fig:Kerr}.
The blue shift of the theoretical results is still present, while the overestimation of the
dipoles in both $\sigma_{xx}(\omega)$ and $\sigma_{xy}(\omega)$ is compensated and
the intensity of the MOKE signal is closer to the experimental data than for the
case of the optical conductivity.

To conclude this section, we have shown that for $Fe$, $Co$ and $Ni$ the Kerr
computed from our pseudo--potential approach are in good agreement with the results obtained
from all--electrons calculations.

\section{Beyond the IP--RPA approximation}\label{sec:MOKE_FULL}
\subsection{A density based approach.}

In the previous section we have constructed the Kerr paramenters starting from the
KS wave--functions using Eq.~\ref{eq:eps_jj}, as it is commonly
done in the literature. However the use of KS wave--function to construct
$\tens{\chi}^0_{\mathbf{jj}}$ is not formally justified. Moreover the inclusion
of LF and xc effects within a density based approach in Eq.~\ref{eq:eps_jj} is
not straightforward.

However, to describe the MOKE only the long wave--length term, i.e. $\mathbf{q}=0$,
of the dielectric function is needed, where the distinction between longitudinal
and transverse fields disappears. In this limit the diagonal part of the dielectric tensor 
can be constructed from $\chi^0_{\rho\rho}$ which,
at finite $\mathbf{q}$, describes only
the longitudinal term of the dielectric tensor. This approach is formally justified within
a density based approach and moreover would allow a straightforward inclusion of LF and xc effects
within the TDDFT scheme.
It is then tempting to try to construct the full dielectric tensor at $\mathbf{q}=0$
from $\chi^0_{\rho\rho}$ and use the result to go beyond the IP--RPA scheme.

Here we provide an heuristic derivation where
only longitudinal fields are considered, as our final goal is to take the
$\vect{q}\rightarrow 0$ limit. We will prove {\it a posteriori} that the result is correct
for systems with an electronic gap or, more in general, when the pure time--reversal
symmetry exist and we will discuss in detail the difference between the derived equation
and Eq.~\ref{eq:eps_jj}.

We consider a non uniform system. The dielectric function is defined as:
\begin{equation} \label{eq:eps1}
\vect{E}^{ext}(\vect{q},\omega)=
\tens{\varepsilon}(\vect{q}\vect{q'},\omega) \vect{E}^{tot}(\vect{q'},\omega)
\text{.}
\end{equation}
Assuming that only longitudinal fields exist, Eq. (\ref{eq:eps1}) can be written in terms of the potentials
\begin{equation} \label{eq:eps2}
V^{ext}(\vect{q},\omega)=
  \vect{\hat{q}}\ \tens{\varepsilon}(\vect{q}\vect{q'},\omega)\ \vect{\hat{q}'}
  V^{tot}(\vect{q'},\omega),
\end{equation}
where the two are related by the equation
\begin{eqnarray}
V^{ext}(\vect{q},\omega)&=& V^{tot}(\vect{q},\omega)-V^{ind}(\vect{q},\omega)= \\
                        &=& V^{tot}(\vect{q},\omega)-
               \frac{4\pi e^2}{q^2}\chi^0_{\rho\rho}(\vect{q},\vect{q'},\omega) V^{tot}(\vect{q'},\omega)
               \label{eq:Vext_def}
\text{.}
\end{eqnarray}

Inserting Eq.~\ref{eq:Vext_def} into Eq.~\ref{eq:eps2} and
taking the limit $\vect{q}\rightarrow \vect{q'} \rightarrow 0$ we can define a generalization of the
relation that holds between $\varepsilon_{\alpha\alpha}$ and $\chi_{\rho\rho}$~\cite{Strinati1988}:
\begin{equation}\label{eq:eps_rr}
\varepsilon_{\alpha\beta}(\omega)=\delta_{\alpha\beta}-\lim_{q_\alpha,q_\beta\rightarrow 0}
                               \frac{4\pi e^2}{q^2} \chi^0_{\rho\rho}(q_\alpha,q_\beta,\omega)
\text{.}
\end{equation}

In order to compare Eq.~\ref{eq:eps_rr} and Eq.~\ref{eq:eps_jj} we first notice that the latter
is divergent for $\omega\rightarrow 0$. After some algebra Eq.~\ref{eq:eps_jj} can be rewritten as~\cite{Sipe1993}:
\begin{equation}\label{eq:eps_jj_w}
\varepsilon_{\alpha,\beta}(\omega)=\frac{A_{\alpha\beta}}{\omega^2}+\frac{B_{\alpha\beta}}{\omega}+\delta_{\alpha\beta}+
          \sum_{cv} \frac{d^3\mathbf{k}}{(2\pi)^3}\ \frac{4\pi e^2\hbar^2}{(\epsilon_{c\mathbf{k}}-\epsilon_{v\mathbf{k}})^2}\
          \[\chi^0_{j_\alpha j_\beta}(\mathbf{0},\omega)\]_{cv\mathbf{k0}}                     
\text{.}
\end{equation}

$A_{\alpha\beta}$ describes the contribution from the electrons at the Fermi surface,
i.e. the Drude term, and is zero in cold semiconductors, when there are not partially filled bands.
This term is also included in Eq.~\ref{eq:eps_rr} in the $q\rightarrow 0$ limit
as discussed in Ref.~\onlinecite{Marini2001}.
Once the $\omega^{-2}$ has been isolated using the relation
$x^\alpha_{cv\mathbf{k}}=-i\hbar p^\alpha_{cv\mathbf{k}}/(m(\epsilon_{c\mathbf{k}}-\epsilon_{v\mathbf{k}}$))
in the last term of Eq.~\ref{eq:eps_jj_w} together with
\begin{eqnarray}
\lim_{\mathbf{q,q'}\rightarrow 0}
\chi^0_{\rho \rho}(\mathbf{q,q'},\omega)       &=& \lim_{\mathbf{q,q'}\rightarrow 0} \sum_{c\ v} \int d^3\mathbf{k} \left [ 
        \frac{(i\mathbf{q}\cdot\mathbf{x}_{cv\mathbf{k}}^*) (i\mathbf{q'}\cdot\mathbf{x}_{cv\mathbf{k}}) \ f_{v\mathbf{k}}(1-f_{c\mathbf{k-q}})}
             {\hbar\omega-(\epsilon_{c\mathbf{k-q}}-\epsilon_{v\mathbf{k}})+i\eta}-  \right .   \nonumber \\
&&\phantom{\chi^0_{\rho \rho}(\mathbf{q},\omega)= \sum_{cv} \int d^3\mathbf{k}} \left .
        \frac{(i\mathbf{q}\cdot\mathbf{x}_{vc\mathbf{k}}^*) (i\mathbf{q'}\cdot\mathbf{x}_{vc\mathbf{k}}) \ f_{v\mathbf{k-1}}(1-f_{c\mathbf{k}})}
             {\hbar\omega+(\epsilon_{c\mathbf{k}}-\epsilon_{v\mathbf{k-q}})+i\eta}
                                                                             \right ],
\end{eqnarray}
we obtain the remaining part of Eq.~\ref{eq:eps_rr}.

\subsection{The anomalous Hall effect}

Hence term $B_{\alpha\beta}$ is not included in Eq.~\ref{eq:eps_rr}.
It can be explicitly written, at the RPA-IP level, as:
\begin{equation}\label{eq:B}
B_{\alpha\beta}=\frac{\hbar e^2}{2\pi^2 m^2} 
\sum_{uw} \int d^3\mathbf{k} (f_{u\mathbf{k}}-f_{w\mathbf{k}})
                          \frac{(p_{wu\mathbf{k}}^\alpha)^*  p_{wu\mathbf{k}}^\beta}{(\epsilon_{w\mathbf{k}}-\epsilon_{u\mathbf{k}})^2}
\text{.}
\end{equation}
This can be shown to be zero when the time--reversal symmetry holds~\cite{Sipe1993} or in any case when 
${\alpha=\beta}$ inverting the mute indexes $u$ and $w$ in the second term on the r.h.s.\ .
In MOKE experiments however the time--reversal is broken by the existence of a ground
state magnetization and by the SOC term in the Hamiltonian and for the construction
of $\Psi_K(\omega)$ we need the terms ${\alpha\neq\beta}$.

Thus $B_{\alpha\beta}$ can differ from zero.
In the following, we briefly discuss its physical meaning.  
To fix the ideas we chose $\alpha=x$ and $\beta=y$. 
It can be easily proved that  the 
the $B_{xy}$ coefficient is (apart a trivial factor raising from the relation between
$\tens{\varepsilon}$ and $\tens{\sigma}$) 
the intrinsic anomalous Hall conductivity (AHC), which is responsible of anomalous
Hall effect in magnetic metals. 

In  fact, according to Ref.~\cite{Yao2004} the AHC reads
\begin{equation}
\sigma^{AHC}_{xy}=
-\frac{e^2}{\hbar}\sum_u\int \frac{d^3 {\bf k}}{(2\pi)^3}
f_{u{\bf k}}  \Omega^z_u({\bf k}); 
\end{equation}
i.e. $\sigma^{AHC}$ can be expressed as a BZ integral of 
the Berry curvature of the $u$-band, 
 $ \Omega^z_u({\bf k})$
(summed over all the occupied states). 
The latter quantity  
can be written  in terms of the ingredients 
of Eq.~\ref{eq:B} as:\cite{Yao2004} 
\begin{equation}
 \Omega^z_u({\bf k})=-\frac{\hbar^2}{m^2} 
\sum_{w,w\ne u} \frac{ 2 Im \left ( p^x_{uw\mathbf{k}} p^y_{wu\mathbf{k}} \right )}
{\left ( \epsilon_{u{\bf k}}- \epsilon_{w{\bf k}} \right )^2 }
\end{equation}

After some straightforward algebra, one can easily prove that 
the AHC can be expressed as 
\begin{equation}
\sigma^{AHC}_{xy}=\frac{B_{xy}}{4\pi i};
\end{equation}
which provide the relations between $B$ and the AHC.  
In the case of magnetic metals, 
our expression constitute an alternative approach to compute $\sigma^{AHC}$ 
with respect to the methods based on the computation of Berry phase~\cite{Wang2007}. 
In the case of insulators instead $\sigma^{AHC}$ has been recognized as a topological invariant~\cite{Thouless1982},
also called Chern number, which can take only integer values
Thus, in dielectric, $B_{xy}$ can be non-zero only in the so-called Chern
insulator, hypothetical materials showing a quantum Hall effect without external magnetic field.
In practice for all the presently known dielectrics
eq.~\ref{eq:eps_rr} can be considered exact.

\begin{figure}[!ht]
\includegraphics[width=0.65\textwidth]{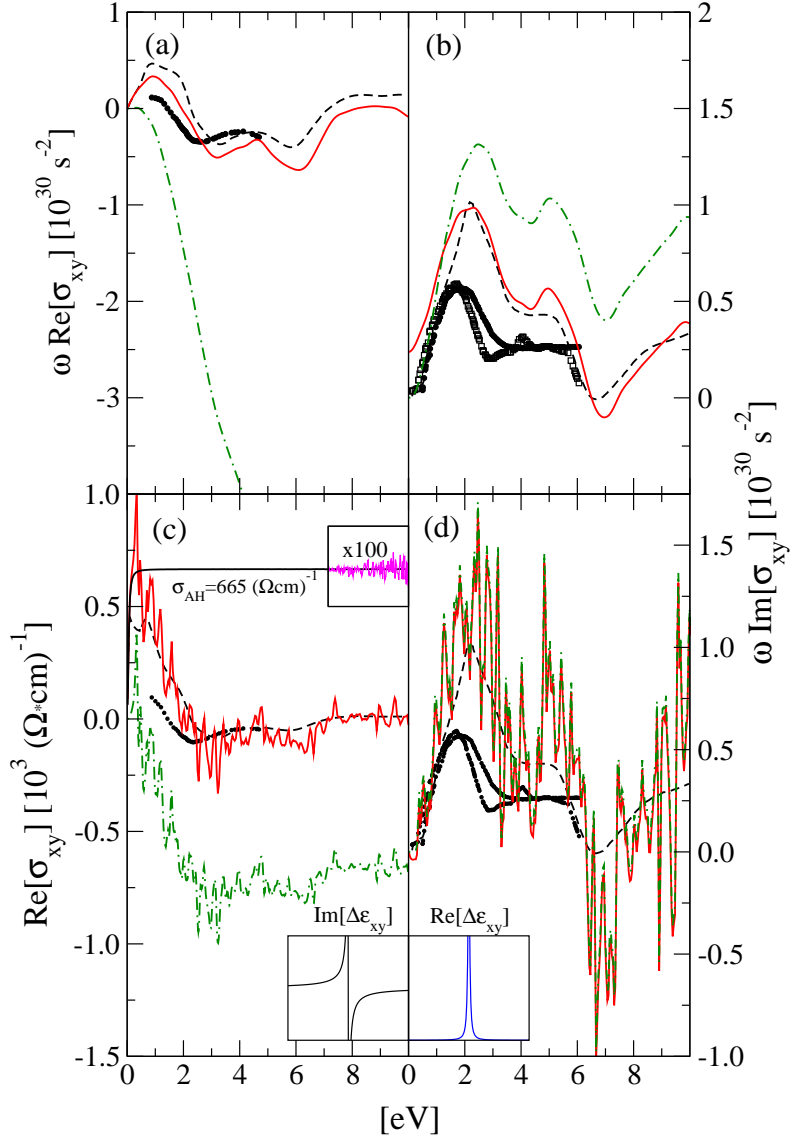}
\caption{(color online) Off--diagonal element of the conductivity tensor, $\sigma_{xy}(\omega)$,
of iron computed starting from Eq.~\ref{eq:eps_rr}, dot--dashed (green) line and
from Eq.~\ref{eq:eps_jj} continuous (red) line.  The dashed line
are all electrons results from Ref.~\onlinecite{Delin1999}. The symbols are experimental
measurements: filled circles, Ref.~\onlinecite{Engen1983}; empty squares,
Ref.~\onlinecite{Krinchik1968}. In panels $a$-$b$ $\sigma_{xy}(\omega)$ is computed with the 
smearing used in the present work. In panels $c$ and $d$ instead $\sigma_{xy}(\omega)$ is evaluated 
at $\eta=0.05\ eV$. The difference between the dot--dashed (green) line and
the continuous (red) line is represented with a thin (black) line. In the inset of the panels
also the difference in terms of the dielectric function is represented.}
\label{fig:anom_Hall}
\end{figure}

Also in this case we have tested, at the IP--RPA level, the effect of $B_{\alpha\beta}$
on bulk iron comparing the conductivity computed
starting either from Eq.~\ref{eq:eps_jj} or from Eq.~\ref{eq:eps_rr}.
In Fig.~\ref{fig:anom_Hall}.$(a-b)$ we show the error induced in the computation of the off-diagonal
conductance on bulk iron.

To clarify the relation between this difference and the AHC also numerically we have considered,
in Fig.~\ref{fig:anom_Hall}.$(c-d)$, the plot of the conductivity at small smearing,
i.e. $\eta=0.05\ eV$, as Eq.\ref{eq:eps_jj} and
Eq.~\ref{eq:eps_rr} are equal only in the limit $\eta\rightarrow 0\ $~\cite{footnote_eta}.
From Fig.~\ref{fig:anom_Hall}.$(c)$ is clear that the difference of the 
two gives a constant value, as expected theoretically, a part from the region $\omega\simeq 0$,
where the $1/\omega^2$ term makes Eq.~\ref{eq:eps_jj} numerically unstable. We can thus extract 
the $\sigma_{xy}^{AHC}=665\ (\Omega cm)^{-1}$ which is not so far from the theoretically computed
value $750.8\ (\Omega cm)^{-1}$ of Ref.~\onlinecite{Yao2004}. The difference is likely due
to the sampling of the BZ. As for the case of the Drude term, the anomalous Hall
conductivity depends on the contribution from the electrons at the Fermi surface
and thus a very fine sampling of the BZ should be needed, which is beyond the scope of
the present work.
Also we see in Fig.~\ref{fig:anom_Hall}.$(d)$ that at small $\eta$ the difference between
the imaginary parts of the conductivity computed with the two approaches goes
to zero (Fig.~\ref{fig:anom_Hall}.$(b)$) as expected from the theoretical derivation.
Finally in the insets we have also represented the difference at the
level of the dielectric function which are $Im[\Delta\varepsilon]\propto 1/\omega$ and
$Re[\Delta\varepsilon]\propto \delta(\omega)$ thus respecting the Krames--Kronig relations.

\subsection{Inclusion of local fields and excitonic effects.}

The generalization of Eq.~\ref{eq:eps_jj} to include LF and $xc$--effects
is non trivial and needs a careful distinction between longitudinal and transverse induced fields.
The result, derived in Ref.~[\onlinecite{DelSole1984}], is to replace the IP
$\tens{\chi}^0_{\mathbf{jj}}$ with the one constructed from the analytical
part of the electron--hole ($eh$) propagator $\overline{L}(12,34)$ solution
of the modified Bethe--Salpeter equation:
\begin{multline}\label{eq:BSEM}
\overline{L}(12,34)=L_0(12,34)+L_0(12,1'2') \\ \Big[\overline{v}(1',3')\delta(1',2')\delta(3',4')
                   -iW(1,2)\delta(1',3')\delta(2',4')\Big] \overline{L}(3'4',34),
\end{multline}
with $1$ representing spatial, time and spin coordinates: $1=(\vect{x_1},t_1,\sigma_1)$.
The long range part of the exchange interaction
${v(1,2)=\delta(t_1-t_2)/|\mathbf{x_1-x_2}|}$ between the electron and the hole is truncated
with the substitution $v\rightarrow\overline{v}$ where in reciprocal space
\begin{equation}
\overline{v}_{\vect{G}}(\vect{q})=
\begin{cases} 4\pi e^2/|\vect{q+G}|^2 & \text{if  $\vect{G\neq 0}$,}
\\0 &\text{if $\vect{G= 0}$.}
\end{cases}.
\end{equation}
From the electron--hole propagator the $\tens{\overline{\chi}}_{\mathbf{jj}}$ and
$\tens{\overline{\chi}}_{\rho\rho}$
are constructed with the relations~\cite{Strinati1988}
\begin{eqnarray}
\overline{\chi}_{\rho\rho}(1,2)   &=&-i\hbar \overline{L}(1,2;1^+,2^+), \label{chi_rr_L}   \\
\tens{\overline{\chi}}_{\mathbf{jj}}(1,2)&=&-i\hbar \frac{-\hbar^2}{4m^2}\[\mathbf{(\nabla_1-\nabla_1')
                          (\nabla_2-\nabla_2')}\overline{L}(1,2;1'2')\]_{1'=1^+,2'=2^+}  \label{chi_jj_L}
\text{,}
\end{eqnarray}
with $1^+=lim_{\tau\rightarrow 0} (\vect{x_1},t_1+\tau,\sigma_1)$
The result is then
\begin{equation}    \label{eq:eps_jj_full} 
\varepsilon_{\alpha,\beta}(\omega)=\(1-\frac{4\pi e^2n}{m\omega^2}\)\delta_{\alpha,\beta}-
                             \frac{4\pi e^2}{\omega^2}\ \overline{\chi}_{j_\alpha j_\beta}(\mathbf{0},\omega)
\text{.}
\end{equation}

A possible strategy to remain within a density based formalism~\cite{footnote_RealTime_DFT},
starting from Eq.~\ref{eq:eps_jj}, could be to use the unphysical $L^{TDDFT}$ replacing
$iW(1,2)\delta(1',3')\delta(2',4')$ with $f_{xc}(1,2)\delta(1',2')\delta(3',4')$ in Eq.~\ref{eq:BSEM}.
However this is not formally justified neither
and at least a current based formalism should be used, i.e.
current--DFT~\cite{Vignale1987,Vignale2004,Abedinpour2010}. 
Indeed for the description of absorption spectra, the diagonal part only of the dielectric function
is commonly constructed from $\overline{\chi}_{\rho\rho}$ to include LF
and xc effects~\cite{footnote_G0} starting from $\varepsilon_{ii}(\omega)[\chi^0_{\rho\rho}]$.
A similar derivation can be used to include LF and xc--effects
replacing  ${\chi^0_{\rho\rho}}$ with ${\overline{\chi}_{\rho\rho}}$ in Eq.~\ref{eq:eps_rr}.
In this case however the Dyson equation for the response function should be written 
for a non homogeneous system assuming, as we did in the IP case, that transverse fields can
be neglected as we are looking for the $\mathbf{q}\rightarrow 0$ limit. The result is:
\begin{equation}\label{eq:eps_rr_full}
\varepsilon_{\alpha\beta}(\omega)=\delta_{\alpha\beta}-\lim_{q_\alpha,q_\beta\rightarrow 0}
                               \frac{4\pi e^2}{q^2} \overline{\chi}_{\rho\rho}(q_\alpha,q_\beta,\omega)
\text{,}
\end{equation}
which, according to the discussion of the previous sections should hold when the time--reversal
symmetry exists or for systems with an electronic gap~\cite{footnote_chern,footnote_future}.
Eq.~\ref{eq:eps_rr_full} must then be compared with Eq.~\ref{eq:eps_jj_full}.
If $\overline{\tens{\chi}}_{\mathbf{jj}}$ and $\overline{\chi}_{\rho\rho}$ are
constructed from the same $\overline{L}$ using Eqs.~\ref{chi_rr_L}-\ref{chi_jj_L} the two are
diagonalized by the same vectors in $c\ v$ space, $A^I_{cv\mathbf{k}}$;
here $I$ is the index of the excitation, which now can be a mixture of electron--hole pairs.
In this case inserting the vectors $A^I_{cv\mathbf{k}}$ in the
equations the two approach will differ by the term
\begin{equation}\label{eq:B_full}
B_{\alpha\beta}=\frac{\hbar e^2}{2\pi^2 m^2} 
\sum_I \sum_{uw} \int d^3\mathbf{k} (f_{u\mathbf{k}}-f_{w\mathbf{k}})
                          \frac{(A^I_{wu\mathbf{k}} p_{wu\mathbf{k}}^\alpha)^* A^I_{wu\mathbf{k}} p_{wu\mathbf{k}}^\beta}
                               {(\hbar\omega_I)^2}
\text{,}
\end{equation}
which defines a generalization of the Anomalous Hall effect.
Here $\omega_I$ are the poles of $\overline{\chi}_{\rho\rho}$.
In common metals usually we have
$A^I_{cv\mathbf{k}}=\delta_{I,(cv)_I}$ (i.e. each vector $A^I$ is different from zero only for a specific
transition $cv=(cv)_I$ ) and $\hbar\omega_I=\epsilon_{ck}-\epsilon_{vk}$, thus
Eq.~\ref{eq:B_full} reduces to Eq.~\ref{eq:B}.

However if one remains within a pure DFT approach, then the vectors which diagonalize $\overline{\chi}_{\rho\rho}$,
$A^{I,TDDFT}_{cv\mathbf{k}}$ do not, in general, diagonalize $L$ and thus $\overline{\tens{\chi}}_{\mathbf{jj}}$.
In this case Eq.~\ref{eq:eps_rr_full} and Eq.~\ref{eq:eps_jj_full} could also differ by a term
proportional to $A^{I,TDDFT}_{cv\mathbf{k}}-A^I_{cv\mathbf{k}}$.
This term must be zero for $\alpha=\beta$, while its relevance in the case $\alpha\neq\beta$
and its eventual physical meaning are left under study.

\section{Conclusions}
We have proposed a scheme to compute the magneto--optical Kerr effect in magnetic--semiconductors.
The scheme has two main novelties. First is based on pseudo--potentials
calculations. This is the most widely used approach to describe extended systems
and we have shown that pseudo--wavefunctions can be used to obtain the Kerr parameters.
The results we find are comparable with all--electrons calculations, provided that the Spin--Orbit
interaction is correctly accounted for in the construction of the pseudo--potential.

Second we have discussed the inclusion of local--field and excitonic effects in the
computation of the MOKE. We have shown that two strategies can be used:
(i) the Bethe--Salpeter equation, through the result derived in Ref.~\onlinecite{DelSole1984},
but also, in almost any case of interest,
(ii) an approach based on time--dependent density--functional
theory and in general on the density--density correlation function
through the result derived in the present manuscript.

\section{Acknowledgments}
This work was partially funded by the Cariplo Foundation through the 
Oxides for Spin Electronic Applications (OSEA) project (n.~2009-2552).
D. Sangalli would like to acknowledge G. Onida and the
European Theoretical Spectroscopy Facility~\cite{ETSF} (ETSF) Milan node for the
opportunity of running simulations on the ETSF--Milano (ETSFMI) cluster,
and P. Salvestrini for technical support on the cluster.
We also acknowledge computational resources provided
by the Consorzio Interuniversitario per le Applicazioni di
Supercalcolo Per Universit\'a e Ricerca (CASPUR) within the project MOSE.
Finally D. Sangalli and A. Debernardi would like to thank R. Colnaghi for technical support.


\begin{thebibliography}{99}
\bibitem{Kerr1877} J. Kerr, Philos. Mag. \textbf{3}, 321 (1877)
\bibitem{Kerr1878} J. Kerr, Philos. Mag. \textbf{5}, 161 (1878)
\bibitem{Bertero1994} G. A. Bertero, and R. Sinclair, J. Magn. Magn. Mater. \textbf{134}, 173 (1994)
\bibitem{Hatwar1997} T. K. Hatwar, Y. S. Tyan, and C. F. Bruker, J. Appl. Phys. \textbf{81}, 3839 (1997)
\bibitem{Alcaraz2012} R. Alcaraz de la Osa, J. M. Saiz, F. Moreno, P. Vavassori, and A. Berger,
                      Phys. Rev. \textbf{B 85}, 064414 (2012)
\bibitem{Bennett1990} W. R. Bennett, W. Schwarzacher, and W. F. Egelhoff, Phys. Rev. Lett. \textbf{65}, 3169 (1990)
\bibitem{Suzuki1998} Y. Suzuki, T. Katayama, P. Bruno, S. Yuasa, and E. Tamura, Phys. Rev. Lett. \textbf{80}, 5200 (1998)
\bibitem{Zinoni2011} C. Zinoni, A. Vanhaverbeke, P. Eib, G. Salis, and R. Allenspach,
                     Phys. Rev. Lett. \textbf{107}, 207204 (2011)
\bibitem{Balk2011} A. L. Balk, M. E. Nowakowski, M. J. Wilson, D. W. Rench, P. Schiffer,
                   D. D. Awschalom, and N. Samarth, Phys. Rev. Lett. \textbf{107}, 077205 (2011)
\bibitem{Wu1999} R. Q. Wu, and A. J. Freeman, J. Magn. Magn. Mater. \textbf{200}, 498 (1999)
\bibitem{Zeper1989} W. B. Zeper, F. J. A. M. Greidanus, P. F. Garcia, and C. R. Fincher,
                    J. Appl. Phys. \textbf{65}, 4971 (1989)
\bibitem{Weller1992} D. Weller, H. Br\"{a}ndle, G. Gorman, C.-J. Lin, and H. Notarys,
                     Appl. Phys. Lett. \textbf{61}, 2726 (1992)
\bibitem{Acbas2009} G. Acbas, M.-H. Kim, M. Cukr, V. Nov\`{a}k, M. A. Scarpulla, O. D. Dubon,
                    T. Jungwirth, Jairo Sinova, and J. Cerne, Phys. Rev. Lett. \textbf{103}, 137201 (2009)
\bibitem{Sun2011} C. Sun, J. Kono, A. Imambekov, and E. Morosan, Phys. Rev. \textbf{B 84}, 224402 (2011)
\bibitem{Yambo} A. Marini, C. Hogan, M. Gr\"uning, and D. Varsano, Comp. Phys. Comm. \textbf{180}, 1392 (2009)
\bibitem{footnote_chern} With  the exception of the so-called Chern insulator.
\bibitem{Guo1995} G. Y. Guo, H. Ebert, Phys. Rev. \textbf{B 51}, 12633 (1995)
\bibitem{Delin1999} A. Delin, O. Eriksson, B. Johansson, S. Auluck, and J. M. Wills, Phys. Rev. \textbf{B 60}, 14105 (1999)
\bibitem{Oppeneer1992} P. M. Oppeneer, T. Maurer, J. Sticht, and J. K\"{u}bler, Phys. Rev. \textbf{B 45}, 10924 (1992)
\bibitem{Oppeneer1995} P. M. Oppeneer, T. Kraft, and H. Eschrig, Phys. Rev. \textbf{B 52}, 3577 (1995)
\bibitem{Kim1999} M. Y. Kim, A. J. Freeman, and R. Wu, Phys. Rev. \textbf{B 59}, 9432 (1999)
\bibitem{Luttinger1967} J. M. Luttinger, \emph{Mathematical Methods in Solid State and Superfluid Theory},
                        edited by R. C. Clark and G. H. Derrick (Oliver and Boyd, Edinburg, 1967), Chap. 4, p. 157
\bibitem{Vernes2002} A. Vernes, L. Szunyogh, P. Weinberger, Phys. Rev. \textbf{B 65}, 144448 (2002)
\bibitem{Vernes2004A} A. Vernes and P. Weinberger, Phys. Rev. \textbf{B 70}, 134411 (2004)
\bibitem{Vernes2004B} A. Vernes, I. Reichl, P. Weinberger, L. Szunyogh, and C. Sommers, Phys. Rev. \textbf{B 70}, 195407 (2004)
\bibitem{Ricci2007} F. Ricci, S. Picozzi, A. Continenza, F. D'Orazio, F. Lucari, K. Westerholt, M. Y. Kim, and A. J. Freeman
                    Phys. Rev. \textbf{B 76}, 014425 (2007)
\bibitem{Stroppa2008} A. Stroppa, S. Picozzi, A. Continenza, M. Y. Kim, and A. J. Freeman,
                      Phys. Rev. \textbf{B 77}, 035208 (2008)
\bibitem{Uba2012} L. Uba, S. Uba, L. P. Germash, L. V. Bekenov, and V. N. Antonov, Phys. Rev. \textbf{B 85}, 125124 (2012)
\bibitem{Ricci2011} Fabio Ricci, Franco D'Orazio, Alessandra Continenza, Franco Lucari, and Arthur J. Freeman,
                    Phys. Rev. \textbf{B 83}, 224421 (2011)
\bibitem{Haidu2011} F. Haidu, M. Fronk, O. D. Gordan, C. Scarlat, G. Salvan, and D. R. T. Zahn,
                    Phys. Rev. \textbf{B 84}, 195203 (2011)
\bibitem{Kubo1957} R. Kubo, J. Phys. Soc. Jpn. \textbf{12}, 570 (1957)
\bibitem{Strinati1988} G. Strinati, Rivista del Nuovo Cimento, Vol. \textbf{11}, 1 (1988)
\bibitem{Read1991} A. J. Read, and R. J. Needs, Phys. Rev. \textbf{B 44}, 13071 (1991)
\bibitem{Johnson1974} P. B. Johnson, and R. W. Christy, Phys. Rev. \textbf{B 9}, 5056 (1974)
\bibitem{Engen1983} P. G. Van Engen, PhD Thesis, Technical University Delft (1983)
\bibitem{Krinchik1968} G. S. Krinchik, and V. A. Artemjev, J. Appl. Phys. \textbf{39}, 1276 (1968)
\bibitem{Shiga1969} M. Shiga, and G. P. Pells, J. Phys. C \textbf{2}, 1847 (1969)
\bibitem{Stoll1971} M. Ph. Stoll, Solid State Comm. \textbf{8}, 1207 (1971)
\bibitem{Ehereinreich1963} H. Ehereinreich, H. R. Philipp, and D. J. Olencha, Phys. Rev. \textbf{131}, 2469 (1963)
\bibitem{Erskine1977} J. L. Erskine, Physica B+C \textbf{89B}, 83 (1977)
\bibitem{Abinit} X. Gonze, B. Amadon, P.-M. Anglade, J.-M. Beuken, F. Bottin, P. Boulanger,
F. Bruneval, D. Caliste, R. Caracas, M. C\^{o}t\`{e}, T. Deutsch, L. Genovese, Ph. Ghosez,
M. Giantomassi, S. Goedecker, D.R. Hamann, P. Hermet, F. Jollet, G. Jomard,
S. Leroux, M. Mancini, S. Mazevet, M.J.T. Oliveira, G. Onida, Y. Pouillon, T. Rangel,
G.-M. Rignanese, D. Sangalli, R. Shaltaf, M. Torrent, M.J. Verstraete, G. Zerah,
J.W. Zwanziger, Comp. Phys. Comm. \textbf{180}, 2582 (2009)
\bibitem{HGH} C. Hartwigsen, S. Goedecker, and J. Hutter, Phys. Rev. \textbf{B 58}, 3641 (1998)
\bibitem{Sipe1993} J. E. Sipe, E. Ghahramani, Phys. Rev. \textbf{B 48}, 11705 (1993)
\bibitem{Virk2007} K. S. Virk, and J. E. Sipe, Phys. Rev. \textbf{B 76}, 035213 (2007)
\bibitem{Giuliani2005} G. F. Giuliani, and G. Vignale, \emph{Quantum Theory of the Electron Fluid},
                   Cambridge University Press, New York (2005); section \textbf{3.4.2} .
\bibitem{footnote_Drude} The inclusion of a big number of $cv$ excitations would be very problematic for the
description of local fields and excitonic effects where a matrix whose dimensions are $N_{cv}xN_{cv}$ must
be diagonalized. 
\bibitem{Marini2001} A. Marini, G. Onida, and R. Del Sole, Phys. Rev. \textbf{B 64}, 195125 (2001)
\bibitem{Katayama1986} T. Katayama, H. Awano, and Y. Nishihara, J. Phys. Soc. Jpn. \textbf{55}, 2539 (1986)
\bibitem{Visnovsky1996} S. Visnovsky, R. Krishnan, M. Nylt, and P. Prosser (1996 - unpublished)
\bibitem{Kawagoe1992} T. Kawagoe, and T. Mizoguchi, J. Magn. Mat. Mater. \textbf{113}, 187 (1992)
\bibitem{Nakajima1996} K. Nakajima, H. Sawada, T. Katayama, and T. Miyazaki, Phys. Rev. \textbf{B 54}, 15950 (1996)
\bibitem{Weller1994} D. Weller, G. R. Harp, R. F. C. Farrow, A. Cebollada, and J. Sticht, Phys. Rev. Lett. \textbf{72}, 2097 (1994)
\bibitem{Yafet2007} J. R. Yates, X. Wang, D. Vanderbilt, and I. Souza,
                   Phys. Rev. \textbf{75}, 195121 (2007)
\bibitem{Wang2006} X. Wang, J. R. Yates, I. Souza, and D. Vanderbilt,
                   Phys. Rev. \textbf{B 74}, 195118 (2006)
\bibitem{Wang2007} X. Wang, D. Vanderbilt, J. R. Yates, and I. Souza, 
                   Phys. Rev. \textbf{B 76}, 195109 (2007)
\bibitem{Yao2004} Y. Yao, L. Kleinman, A. H. MacDonald, J. Sinova, T. Jungwirth, D.S. Wang, E. Wang, and Q. Niu,
                  Phys. Rev. Lett. \textbf{92}, 037204 (2004)
\bibitem{Thouless1982} D. J. Thouless, M. Kohmoto, M. P. Nightingale, and M. den Nijs
                       Phys. Rev. Lett. \textbf{49}, 405 (1982)
\bibitem{footnote_eta} In particular the expression $(\omega+i\eta)^{-1}$ is used 
in the implementation which in the limit $\eta\rightarrow 0$ gives
\begin{equation*}
\lim_{\eta\rightarrow 0}\frac{1}{\omega+i\eta}=\frac{1}{\omega}-i\pi \delta(\omega)
\text{.}
\end{equation*}
\bibitem{DelSole1984} R. Del Sole, and E. Fiorino, Phys. Rev. \textbf{B 29}, 4631 (1984)
\bibitem{footnote_RealTime_DFT} For completeness we also mention that an alternative approach to the inclusion of
LF and xc--effects within TDDFT can be based on the real--time propagation approach~\cite{Varsano2009}.
\bibitem{Vignale1987} G. Vignale and M. Rasolt, Phys. Rev. Lett. \textbf{59}, 2360 (1987)
\bibitem{Vignale2004} G. Vignale, Phys. Rev. \textbf{B 70}, 201102(R) (2004)
\bibitem{Abedinpour2010} S. H. Abedinpour, G. Vignale, and I. V. Tokatly Phys. Rev. \textbf{B 81}, 125123 (2010)
\bibitem{footnote_G0} The analytic part only of the electron--hole propagator
is used to better describe the long range part of the coulomb interaction in
extended systems.
Indeed, within the dipole approximation, it is also possible to express $\varepsilon^{-1}(\omega)[\chi_{\rho\rho}]$.
This is equivalent to $\varepsilon(\omega)[\overline{\chi}_{\rho\rho}]$.
The substitution $\varepsilon^{-1}(\omega)[\chi_{\rho\rho}]\rightarrow \varepsilon^{-1}(\omega)[\chi^0_{\rho\rho}]$ leads
to the IP approximation while the replacement
$\varepsilon(\omega)[\overline{\chi}_{\rho\rho}]\rightarrow \varepsilon(\omega)[\chi^0_{\rho\rho}]$ leads to the
IP--RPA approxiamtion where the long range contribution of the Coulomb interaction is described at the 
RPA level, while microscopic fields are described at the IP level, i.e. are neglected.
\bibitem{footnote_future} We are presently doing simulations on magnetic semiconductors, namely $EuS$ and $EuO$,
to check the inclusion of LF and xc effects on the Kerr parameters.
\bibitem{ETSF} A. Y. Matsuura, N. Thrupp, X. Gonze, Y. Pouillon, G. Bruant, and G. Onida, Comput. Sci. Eng. 14, 22 (2012).
(http://www.etsf.it; http://www.etsf.eu)
\bibitem{Varsano2009}  D. Varsano, L. A. Espinosa-Leal, X. Andrade, M. A. L. Marques, R. Di Felice, and A. Rubio,
                       Phys. Chem. Chem. Phys. \textbf{11}, 4481 (2009)
\end{thebibliography}
\end{document}